\magnification 1040
\baselineskip 15 pt

\def \P {{\bf P}}
\def \E {{\bf E}}

\def \D {{\bf D}}

\def \Do {{\cal R}}
\def \bDo {\bar {\cal R}}
\def \la {\lambda}

\def \noi {\noindent}

\def \phis {\phi^s}
\def \Phis {\Phi^s}

\def \bt {\bar \theta}
\def \bphi {\bar \phi}
\def \bw {\bar w}
\def \bwt {\bar w^t}
\def \bE {\bar E}

\def \bphit {\bar \phi^t}
\def \bPhit {\bar \Phi^t}

\def \bPhi {\bar \Phi}

\input srctex.sty

\def \sect#1{\bigskip  \noindent{\bf #1} \medskip }
\def \subsect#1{\bigskip \noindent{\it #1} \medskip}
\def \thm#1#2{\medskip \noindent {\bf Theorem #1.}   \it #2 \rm \medskip}
\def \prop#1#2{\medskip \noindent {\bf Proposition #1.}   \it #2 \rm \medskip}
\def \cor#1#2{\medskip \noindent {\bf Corollary #1.}   \it #2 \rm \medskip}
\def \pf {\noindent  {\it Proof}.\quad }
\def \lem#1#2{\medskip \noindent {\bf Lemma #1.}   \it #2 \rm \medskip}
\def \ex#1{\medskip \noindent {\bf Example #1.}}

\def \rem#1{\medskip \noindent {\bf Remark #1.}}

\def\sqr#1#2{{\vcenter{\vbox{\hrule height.#2pt\hbox{\vrule width.#2pt height#1pt \kern#1pt\vrule width.#2pt}\hrule height.#2pt}}}}

\def \square{\hfill\mathchoice\sqr56\sqr56\sqr{4.1}5\sqr{3.5}5}

\def \qed {$\square$ \medskip}

\nopagenumbers

\headline={\ifnum\pageno=1 \hfill \else \hfill {\rm \folio} \fi}

\centerline{\bf Purchasing Life Insurance to Reach a Bequest Goal}

\bigskip

\centerline{Erhan Bayraktar}
\centerline{Department of Mathematics, University of Michigan}
\centerline{Ann Arbor, Michigan, USA, 48109} \bigskip

\centerline{S. David Promislow}
\centerline{Department of Mathematics, York University}
\centerline{Toronto, Ontario, Canada, M3J 1P3} \bigskip

\centerline{Virginia R. Young}
\centerline{Department of Mathematics, University of Michigan}
\centerline{Ann Arbor, Michigan, USA, 48109} \bigskip


\noindent{\bf Abstract:}  We determine how an individual can use life insurance to meet a bequest goal.  We assume that the individual's consumption is met by an income from a job, pension, life annuity, or Social Security.  Then, we  consider the wealth that the individual wants to devote towards heirs (separate from any wealth related to the afore-mentioned income) and find the optimal strategy for buying life insurance to maximize the probability of reaching a given bequest goal.  We consider life insurance purchased by a single premium, with and without cash value available.   We also consider irreversible and reversible life insurance purchased by a continuously paid premium; one can view the latter as (instantaneous) term life insurance.

\bigskip

\noindent{\it Keywords:} Term life insurance, whole life insurance, bequest motive, deterministic control.

\sect{1. Introduction}

Life insurance helps in estate planning, specifically, in providing bequests for children, grandchildren, or charitable organizations.  With this purpose in mind, we determine how an individual can use life insurance to meet a bequest goal.  We assume that the individual's consumption is met by an income from a job, pension, life annuity, or Social Security.  Then, we consider the wealth that the individual wants to devote towards heirs (separate from any wealth related to the afore-mentioned income) and find the optimal strategy for buying life insurance to maximize the probability of reaching a given bequest goal.

In this paper, we join two hitherto unconnected streams of literature.  The first stream is that of optimal purchasing of life insurance, and most of the research in this area maximizes utility of consumption, bequest, or both.  The seminal article in this area is Richard (1975); please see Bayraktar and Young (2013) for some recent references relevant to the problem of maximizing utility of household consumption by using life insurance.

The second stream is that of maximizing the probability of reaching a particular target.  This problem has been studied in probability problems related to gambling, as in Dubins and Savage (1965, 1976).  For an important extension of the work of Dubins and Savage, see Pestien and Sudderth (1985), in which they control a diffusion process to reach a target before ruining.  For related papers, see Sudderth and Weerasinghe (1989), Kulldorff (1993), and Browne (1997, 1999a, 1999b).  Instead of controlling a diffusion, we maximize the probability of reaching a particular goal and allow the individual to purchase life insurance to help reach that goal, while adding a random deadline (namely, death).

The rest of the paper is organized as follows: In Section 2.1, we consider the case for which the individual buys whole life insurance via a single premium with no cash value available, while in Section 2.2, she can surrender any or all of her whole life insurance for a cash value.  In both cases, we compute her expected wealth at death because her goal is to reach a given bequest, so expected wealth at death is relevant.  Section 3 parallels Section 2 for the case in which insurance is purchased via a continuously-paid premium; however, we reverse the order of the topics as compared with the order in Section 2.  In Section 3.1, the individual is allowed to change the amount of her insurance at any time; in our time-homogeneous setting, this amounts to instantaneous term life insurance.  By contrast, in Section 3.2, we do not allow the individual to terminate life insurance, so for the remainder of her life, she has to pay for any life insurance she buys.  The solution of th!
 e problem in Section 3.1 is simpler than and informs the solution to the problem in Section 3.2, so we present the simpler problem first.  Section 4 concludes the paper.

\sect{2.  Single-Premium Life Insurance}

We begin this section by stating the optimization problem that the individual faces.  In Section 2.1, we consider the case for which the individual buys whole life insurance via a single premium with no cash value available, so she never surrenders her life insurance policy; she may only buy more.  In Section 2.2, we incorporate a non-zero cash value and find the optimal insurance purchasing and surrendering policies in that case.  At the end of each of Sections 2.1 and 2.2, we compute her expected wealth at death.

\subsect{2.1. No cash value available}

We assume that the individual has an investment account that she uses to reach a given bequest goal $b$.  This account is separate from the money that she uses to cover her living expenses.  The individual may invest in a riskless asset earning interest at the continuous rate $r > 0$, which actuaries call the {\it force of interest}, or she may purchase whole life insurance. 

Denote the future lifetime random variable of the individual by $\tau_d$.  We assume that  $\tau_d$ follows an exponential distribution with mean ${1 \over \la}$.  (In other words, the individual is subject to a constant {\it force of mortality}, or hazard rate, $\la$.)  The individual buys life insurance that pays at time $\tau_d$.  This insurance acts as a means for achieving the bequest motive.  In this time-homogeneous model, we assume that a dollar death benefit payable at time $\tau_d$ costs $H$ at any time.  Write the single premium as follows:
$$
H = (1 + \theta) \bar A_{x} = (1 + \theta) {\la \over r+ \la},
$$
in which $\theta \ge 0$ is the proportional risk loading.  Assume that $\theta$ is small enough so that $H < 1$; otherwise, if $H \ge 1$, then the buyer would not pay a dollar or more for one dollar of death benefit. 

In this section and in Section 2.2, we suppose that the premium is payable at the moment of the contract; as stated above, $H$ is the single premium per dollar of death benefit.  In Section 3, we consider the case for which the insurance premium is payable continuously.

Let $W(t)$ denote the wealth in this separate investment account at time $t \ge 0$.  Let $D(t)$ denote the amount of death benefit payable at time $\tau_d$ purchased at or before time $t \ge 0$.   Thus, with single-premium life insurance, wealth follows the dynamics
$$
\left\{
\eqalign{
dW(t) &= r W(t-) \, dt - H \, dD(t), \quad 0 \le t < \tau_d, \cr
W(\tau_d) &= W(\tau_d-) + D(\tau_d-).
}
\right.
$$

An insurance purchasing strategy $\D = \{  D(t) \}_{t \ge 0}$ is {\it admissible} if (i) $\D$ is a non-negative, non-decreasing process, independent of $\tau_d$, and (ii) if wealth under this process is non-negative for all $t \ge 0$.  We include the latter condition to prevent the individual from borrowing against her life insurance.

\rem{2.1} {By requiring that $\D$ be non-decreasing over time, we effectively assume the individual cannot surrender any life insurance once she has bought it.  In the real world, whole life insurance has a surrender value that the individual can withdraw, and in Section 2.2, we include that feature. \qed}

We assume that the individual seeks to maximize the probability that $W(\tau_d) \ge b$, by optimizing over admissible controls $\D$.  The corresponding value function is given by
$$
\phi(w, D) = \sup_{\D} \P^{w, D} \left( W(\tau_d) \ge b \right),
\eqno(2.1)
$$
in which $\P^{w, D}$ denotes conditional probability given $W(0-) = w \ge 0$ and $D(0-) = D \ge 0$.  We call $\phi$ the maximum probability of reaching the bequest goal.

If $D \ge b$, then the individual has already reached her bequest goal of $b$; thus, henceforth, in this section, we assume that $D < b$.  If wealth equals $H(b - D)$, the so-called {\it safe level}, then it is optimal for the individual to spend all of her wealth to purchase life insurance of $b - D$ so that her total death benefit becomes $b = (b - D) + D$.  It follows that $\phi(w, D) = 1$ for $w \ge H(b - D)$ and $0 \le D < b$.  Thus, it remains only to determine the maximum probability of reaching the bequest on $\Do = \{ (w, D): 0 \le w \le H(b - D), \, 0 \le D < b \}$.

We next present a verification lemma that states that a ``nice'' solution to a variational inequality associated with the maximization problem in (2.1) is the value function $\phi$.  Therefore, we can reduce our problem to one of solving a variational inequality.  We state the verification lemma without proof because its proof is similar to others in the literature; see, for example, Wang and Young (2012a, 2012b) for related proofs in a financial market that includes a risky asset.

\lem{2.1} {Let $\Phi = \Phi(w, D)$ be a function that is non-decreasing and differentiable with respect to both $w$ and $D$  on $\Do = \{ (w, D): 0 \le w \le H(b - D), \, 0 \le D < b \}$, except that $\Phi$ might have infinite derivative with respect to $w$ at $w = 0$.  Suppose $\Phi$ satisfies the following variational inequality on $\Do$, except possibly when $w = 0$:
$$
\max( rw \, \Phi_w - \la \, \Phi_w, \, \Phi_D - H \, \Phi_w) = 0.
\eqno(2.2)
$$
\noi Additionally, suppose $\Phi(H(b-D), D) = 1$. Then, on $\Do$,
$$
\phi = \Phi.
\eqno{\square}
$$}

The region ${\cal R}_1 = \{ (w, D) \in \Do: \phi_D(w, D) - H \, \phi_w(w, D) < 0 \}$ is called the {\it continuation} region because when the wealth and life insurance benefit lie in the interior of ${\cal R}_1$, the individual does not purchase additional life insurance; rather, she {\it continues} with her current benefit and invests her wealth in the riskless asset.  Indeed, $\phi_D < H \, \phi_w$ means that the marginal benefit of buying more life insurance ($\phi_D$) is less than the marginal cost of doing so ($H \, \phi_w$).  On the closure of that region in $\Do$, written $cl({\cal R}_1)$, the equation $\la \phi = rw \phi_w$ holds.

To help us solve the variational inequality (2.2), we recall that in similar problems (for example, purchasing life annuities to minimize the probability of lifetime ruin, as described in Milevsky et al.\ (2006)), the optimal strategy is to ``act'' only at the safe level.  In our case, that translates into buying life insurance only when wealth reaches $H(b - D)$, so that $\phi$ solves the following boundary-value problem for $0 \le w \le H(b - D)$ and $0 \le D < b$:
$$
\left\{
\eqalign{
& r w  \phi_w - \la \phi  = 0, \cr
& \phi(H(b - D), D) = 1.
} \right.
\eqno(2.3)
$$
Buying life insurance only when wealth reaches $H(b - D)$ is indeed optimal, as we prove in the following proposition.

\prop{2.2} {The maximum probability of reaching the bequest goal on $\Do = \{ (w, D): 0 \le w \le H(b - D), \, 0 \le D < b \}$ is given by
$$
\phi(w, D) = \left( {w \over H(b - D)} \right)^{{\la \over r}}.
\eqno(2.4)
$$
The associated optimal life insurance purchasing strategy is not to purchase additional life insurance until wealth reaches the safe level $H(b - D)$, at which time, it is optimal to buy additional life insurance of $b - D$.}

\pf We use Lemma 2.1 to prove this proposition.  First, note that $\phi$ in (2.4) is increasing and differentiable with respect to both $w$ and  $D$ on $\Do$, except possibly at $w = 0$.  Because $\phi$ solves the boundary-value problem (2.3), we have $rw \phi_w - \la \phi = 0$ on $\Do$.

Next, we show that $\phi_D - H \, \phi_w \le 0$ on $\Do$:
$$
\phi_D(w, D) - H \, \phi_w(w, D) =  {\la \over rH} \left( {w \over H(b - D)} \right)^{{\la \over r} -1} \left[ {w \over (b - D)^2} - {H \over b - D} \right] \propto w - H(b - D) \le 0.
$$
We have, thus, shown that the expression for $\phi$ in (2.4) satisfies the variational inequality (2.2).

The continuation region equals $\Do_1 = \{(w, D): 0 \le w < H(b - D), \, D < b \}$; therefore, the optimal insurance purchasing strategy is to buy additional insurance of $b - D$ when wealth reaches the safe level $H(b - D)$. \qed

\ex{2.1}   For  a  numerical   example   take the following parameter values:  $b = 1$, $r = 0.03$, $\la = 0.08$, $ \theta = 0.2$.    Suppose  a  person  has    wealth of   0.4 , and    carries existing  insurance of  0.3.   The optimal strategy is to wait until    wealth reaches  the safe level  of    0.60909 and then  buy.  The probability  that   the  bequest goal  is  achieved  is        0.11198.

\rem{2.2} {We fully anticipate that the results of this section will hold when one considers other models, such as more general financial and mortality models, including those that are not time homogeneous. Specifically, we expect that when insurance is purchased by a single premium with no cash value available, then it will be optimal to wait until wealth reaches the safe level to buy additional life insurance.}

\rem{2.3} {Optimally controlled wealth is invested in the riskless asset until it reaches $H(b-D)$; thus, wealth at time $t$, before reaching the safe level, equals $W(t) = w e^{rt}$, for a given initial wealth $w < H(b-D)$.  The time that wealth reaches the safe level, denoted by $\tau_{H(b-D)}$, is given by
$$
\tau_{H(b-D)} = {1 \over r} \, \ln \left( {H(b-D) \over w} \right).
$$
The individual reaches her bequest motive if she dies after time $\tau_{H(b-D)}$; this occurs with probability $e^{-\la \tau_{H(b-D)}}$, which equals the expression given in (2.4), as expected. \qed}

\smallskip

Because we are maximizing the probability that wealth at death equals $b$, it is of interest to determine the expected wealth at death.  Expected wealth at death equals $E(w, D) = \E^{w, D} (W(\tau_d-) + D(\tau_d-))$; thus, from the discussion in Remark 2.3, we have
$$
E(w, D) = \int_0^{\tau_{H(b-D)}} (w e^{rt} + D) \, \la e^{-\la t} \, dt + b \, \phi(w, D),
$$
and the corollary below follows.

\cor{2.3} {Expected wealth at death, $E(w, D) = \E^{w, D}(W(\tau_d))$, for an individual who follows the optimal life insurance purchasing strategy of Proposition $2.2$ is given by
$$
E(w, D) =
\cases{ (b-D) \left[ 1 - {\la H \over \la - r} \right] \left( {w \over H(b-D)} \right)^{{\la \over r}} + {\la w \over \la - r} + D, &if $\la \ne r$, \cr \cr
w \left[ {1 \over H} + \ln \left( {H(b-D) \over w} \right) \right] + D, &if $\la = r$.
}
\eqno(2.5)
$$}

\rem{2.4} {An expectation such as $E^{w,D}(W(\tau_d))$ satisfies a differential equation with boundary conditions.  Indeed, via a standard verification lemma, one can show that $\E^{w,D}(W(\tau_d))$ uniquely solves the following boundary-value problem (BVP) for $0 \le w \le H(b-D)$ and $0 \le D < b$:
$$
\left\{
\eqalign{
& \la (E - (w+D)) =  r w \, E_w, \cr
& E(H(b-D), D) = b.
}
\right.
$$
Because the expression in equation (2.5) solves this BVP, we confirm that it is the correct expression for $\E^{w,D}(W(\tau_d))$.}

\subsect{2.2.  Cash value available}

Standard nonforfeiture laws ensure that an individual who owns a whole life insurance policy can exchange the policy for its cash value.  In this section, we incorporate that feature of whole life insurance into the model in Section 2.1.  Therefore, we allow the process $\D$ to decrease, although it is still required to be non-negative.

We assume that when the individual surrenders her death benefit, she receives a proportion of the purchase price.  (If the cash value is determined according to some other method, such as a proportion of the reserve, then one can still express it as a proportion of the purchase price.)  Let $\rho \in [0, 1]$ be the proportional surrender charge, so that the individual receives $(1 - \rho) H$ for each dollar of death benefit that she surrenders.  The case in which $\rho = 1$ is equivalent to the case for which no cash value is available, as in Section 2.1.

Write $\phis$ for the maximum probability of wealth at death reaching the bequest $b$ when whole life insurance can be  surrendered.  (We use a superscript $s$ to denote that insurance can be surrendered.)  The corresponding verification lemma is as follows.

\lem{2.4} {Let $\Phis = \Phis(w, D)$ be a function that is non-decreasing, continuous, and piecewise differentiable with respect to both $w$ and $D$ on $\Do = \{ (w, D): 0 \le w \le H(b - D), \, 0 \le D < b \}$.  Suppose $\Phis$ satisfies the following variational inequality on $\Do$:
$$
\max( rw \, \Phis_w - \la \, \Phis, \, \Phis_D - H \, \Phis_w, \, (1 - \rho) H \, \Phis_w - \Phis_D) = 0,
\eqno(2.6)
$$
in which we use one-sided derivatives, if needed. Additionally, suppose $\Phis(H(b-D), D) = 1$. Then, on $\Do$,
$$
\phis = \Phis.
\eqno{\square}
$$}

The region ${\cal R}_1 = \{ (w, D) \in \Do: \phis_D(w, D) - H \, \phis_w(w, D) < 0 \hbox{ and } (1 - \rho) H \, \phis_w(w, D) - \phis_D(w, D) < 0 \}$ is the {\it continuation} region because when the wealth and life insurance benefit lie in the interior of ${\cal R}_1$, the individual does not purchase nor surrender life insurance; she {\it continues} with her current benefit.  After Lemma 2.1, we discussed the inequality $\phis_D < H \, \phis_w$; to review, it means that the marginal benefit of buying more life insurance ($\phis_D$) is less than the marginal cost of doing so ($H \, \phis_w$).  Similarly, inequality $(1 - \rho) H \, \phis_w < \phis_D$ means that the marginal benefit of surrendering life insurance $\left( (1 - \rho) H \, \phis_w \right)$ is less than the marginal cost of doing so ($\phis_D$).  On the closure of that region in $\Do$, written $cl({\cal R}_1)$, the equation $\la \phis = rw \phis_w$ holds.

To find $\phis$, we hypothesize that the optimal purchasing strategy is identical to the one in Section 2.1.  Specifically, the individual does not buy additional insurance until wealth reaches the safe level $H(b - D)$.  Furthermore, we hypothesize that it is optimal to surrender life insurance for wealth small enough, so that the individual liquidates her assets in order to take advantage of the riskless return.  It turns out that this hypothesis is correct, and we prove this assertion in the following proposition.

\prop{2.5} {The maximum probability of reaching the bequest goal on $\Do = \{ (w, D): 0 \le w \le H(b - D), \, 0 \le D < b \}$ is given by
$$
\phis(w, D) = 
\cases{
\left( {w + (1-\rho) H D \over Hb} \right)^{\la \over r}, &if $0 \le w < (1-\rho) H(b - D) $, \cr \cr
\left( {w \over H(b - D)} \right)^{\la \over r}, &if $(1-\rho) H(b - D) \le w \le H(b - D)$.
}
\eqno(2.7)
$$
The associated optimal life insurance surrendering and purchasing strategies are as follows:
\item{$(a)$} If wealth is less than $(1-\rho) H(b - D),$ then surrender {\it all} life insurance.  Thereafter, invest all wealth in the riskless asset until wealth reaches the safe level $Hb$, at which time, it is optimal to buy life insurance of $b$.
\item{$(b)$} If wealth is greater than or equal to $(1-\rho)H(b-D)$, then invest all wealth in the riskless asset until wealth reaches the safe level $H(b - D)$, at which time, it is optimal to buy additional life insurance of $b-D$.
}

\pf We use Lemma 2.4 to prove this proposition.  First, note that $\phis$ in (2.7) is increasing, continuous, and piecewise differentiable with respect to both $w$ and $D$ on $\Do$.  When $0 \le w < (1-\rho)H(b - D)$,
$$
rw \, \phis_w - \la \phis = - {\la (1-\rho)D \over b} \left( {w + (1-\rho)HD \over Hb}  \right)^{{\la \over r}-1} \le 0.
$$
In fact, this inequality holds strictly, except when $D = 0$, in which case the individual has no death benefit to surrender.  For $(1-\rho)H(b - D) \le w \le H(b - D)$, $\phis$ solves the differential equation in (2.3); thus, $rw \, \phis_w - \la \, \phis \le 0$ on $\Do$.

Next, observe that $(1-\rho)H \, \phis_w - \phis_D \le 0$ on $\Do$.  Indeed, when $0 \le w < (1-\rho)H(b - D)$, the inequality holds with equality, while $rw \, \phis_w - \la \phis < 0$ for $D \ne 0$; thus, it is optimal to surrender all one's life insurance when wealth is less than $(1-\rho)H(b - D)$.  When $(1-\rho)H(b - D) < w \le H(b - D)$, the inequality holds strictly; thus, we deduce that it is not optimal to surrender any life insurance when wealth is greater than $(1-\rho)H(b - D)$.

When $w = (1 - \rho)H(b-D)$, the individual is indifferent between surrendering all her life insurance and surrendering none of it, as far as maximizing the probability that she will die with wealth equal to $b$.  We assume that she surrenders none of her life insurance when $w = (1-\rho)H(b-D)$ because, for that wealth, expected wealth at death is greater when she does not surrender her life insurance; see Corollary 2.6 below.

Finally, observe that $\phis_D(w, D) - H \, \phis_w(w, D) \le 0$ on $\Do$.  Indeed, when $0 \le w < H(b-D)$, the inequality holds strictly; thus, we deduce that it is not optimal to buy additional life insurance until wealth reaches the safe level.  We have, thus, shown that the expression for $\phis$ in (2.7) satisfies the variational inequality (2.6).   \qed


\rem{2.5} {We anticipate that the results of this section will hold when one considers other models, such as more general financial and mortality models, including those that are not time homogeneous. Specifically, we expect that when insurance is purchased by a single premium with cash value available, then it will be optimal to wait until wealth reaches the safe level to buy additional life insurance, and it will be optimal to surrender life insurance when wealth is low enough.  \qed}

\ex{2.2}   Consider again  the  data  of  Example 2.1, only now suppose that  surrender  is allowed.
   If  $\rho = 0.5$, the  success probability of the optimal strategy  rises  to  0.24487.   If  $\rho = 0.3   $  it  rises  further  to    0.31703. \bigskip

As in Section 2.1, it is of interest to determine the expected wealth at death for someone who is allowed to surrender her life insurance in exchange for its cash value.  Computing the expressions in the following corollary is straightforward, so we leave that work for the interested reader.

\cor{2.6} {Expected wealth at death, $E^s(w, D) = \E^{w, D}(W(\tau_d))$, for an individual who follows the optimal life insurance purchasing and surrendering strategies of Proposition $2.5$ is given by the following if $\la \ne r$:
$$
E^s(w, D) =
\cases{
b \left[ 1 - {\la H \over \la - r} \right] \left( {w + (1- \rho)HD \over Hb} \right)^{{\la \over r}} + {\la (w + (1- \rho)HD) \over \la - r}, &if $0 \le w < (1-\rho)H(b-D)$, \cr \cr
(b-D) \left[ 1 - {\la H \over \la - r} \right] \left( {w \over H(b-D)} \right)^{{\la \over r}} + {\la w \over \la - r} + D, &if $(1-\rho)H(b-D) \le w \le H(b-D)$.
}
\eqno(2.8)
$$
If $\la = r$, then expected wealth at death is given by
$$
E^s(w, D) =
\cases{
(w + (1- \rho)HD) \left[ {1 \over H} + \ln \left( {Hb \over w + (1- \rho)HD} \right) \right], &if $0 \le w < (1-\rho)H(b-D)$, \cr \cr
w \left[ {1 \over H} + \ln \left( {H(b-D) \over w} \right) \right] + D, &if $(1-\rho)H(b-D) \le w \le H(b-D)$.
}
\eqno(2.9)
$$}


\rem{2.6} {Note that $E^s$ in (2.8) and (2.9) is not continuous at $w = (1-\rho)H(b-D)$, which is due to the difference between the optimal surrendering strategy of the individual for wealth less than versus greater than $(1-\rho)H(b-D)$.  One can show that $E^s((1-\rho)H(b-D)-, D) \le E^s((1-\rho)H(b-D)+, D)$; thus, from the standpoint of expected wealth at death, it is better for the individual not to surrender her life insurance when $w = (1-\rho)H(b-D)$, even though the probability of reaching $b$ is the same whether she surrenders all her life insurance or surrenders none at that level of wealth.}

\sect{3. Insurance Purchased by a Continuously Paid Premium}

Section 3 parallels Section 2 for the case in which insurance is purchased via a continuously-paid premium; however, we reverse the order of the subsections.  In Section 3.1, the individual is allowed to change the amount of her insurance at any time; in our time-homogeneous setting, this amounts to instantaneous term life insurance.  By contrast, in Section 3.2, we do not allow the individual to terminate life insurance, so for the remainder of her life, she has to pay for any life insurance she buys.  The solution of the problem in Section 3.1 is simpler than and informs the solution to the problem in Section 3.2, so we present the simpler problem first.

\subsect{3.1  Instantaneous term life}

In this section, we assume that the individual buys life insurance via a premium paid continuously at the rate of $h = (1 + \bt) \la$ per dollar of insurance for some $\bt \ge 0$.   Furthermore, we assume that the individual can change the amount of her insurance coverage at any time.  The proportional loading covers expenses, profit, and risk margin; therefore, we assume that no reserve accumulates.  Thus, the set up in this section is equivalent to the individual purchasing instantaneous term life insurance.  With continuously paid premium for instantaneous term life insurance, wealth follows the dynamics
$$
\left\{
\eqalign{
dW(t) &= (r W(t) - h D(t)) \, dt, \quad 0 \le t < \tau_d, \cr
W(\tau_d) &= W(\tau_d-) + D(\tau_d-) \, .
}
\right.
\eqno(3.1)
$$

For this section, an {\it admissible} insurance strategy $\D = \{ D(t) \}_{t \ge 0}$ is any non-negative process.  We do not insist admissible strategies be such that $W(t) \ge 0$ for all $t \ge 0$ with probability one because of the constant drain on wealth by the negative drift term $- h D(t)$. Therefore, we modify the definition of the maximized probability of reaching the bequest by effectively ending the game if wealth reaches 0 before the individual dies.  Define $\tau_0 = \inf \{ t \ge 0: W(t) \le 0 \}$, and define the value function by
$$
\bphit(w) = \sup_{\D} \P^{w} \left( W(\tau_d \wedge \tau_0) \ge b \right),
\eqno(3.2)
$$
in which we maximize over admissible strategies $\D$.  (We use a {\it bar} to denote that the premium is payable continuously, and we use a superscript $t$ to indicate that the insurance is term life.)  We refer to $\bphit$ as the maximum probability of reaching the bequest goal before ruining.

To motivate the verification lemma for this problem, we present the following informal discussion.  Because $\D$ is an instantaneous control, we anticipate that $\bphit$ solves the following control equation:
$$
\la \, \bphit = rw \, \bphit_w +  \max_D \left[ \la {\bf 1}_{\{w + D \ge b\}} - hD \, \bphit_w  \right],
\eqno(3.3)
$$
in which the indicator function ${\bf 1}_{\{w + D \ge b\}}$ equals 1 if $w + D \ge b$ and equals 0 otherwise.  We did not encounter this indicator function in the problem in Section 2 because for $0 \le w \le H(b - D)$ and $0 \le D < b$, we automatically have $w + D < b$.  

In (3.3), the indicator function equals 0 or 1, and corresponding to each of those values, we choose $D$ to be a (constrained) minimum because of the term $-hD \bphit_w$, that is, constrained by either $w + D < b$ or $w + D \ge b$, respectively.  Specifically, if the indicator equals 0, then the optimal insurance is $D = 0$; if it equals 1, then the optimal insurance is $D = b-w$.  Thus, we can replace equation (3.3) with the equivalent expression:
$$
\la \, \bphit = rw \, \bphit_w +  \max \left[ \la - h (b-w) \, \bphit_w, \, 0  \right].
$$

Denote the safe level for this problem by $\bwt$.  We obtain $\bwt$ by arguing as follows:  the income $r \bwt$ can fund a death benefit of ${r \bwt \over h}$, and we require that sum of this death benefit and the existing wealth $\bwt$ equals the goal $b$; that is, ${r \bwt \over h} + \bwt = b$, or equivalently, $\bwt = {hb \over r + h}$.  Note that $\bwt$ equals the single premium to buy whole life insurance of $b$ when the hazard rate used in pricing is $h$.

These observations lead to the following verification lemma.

\lem{3.1} {Let $\bPhit = \bPhit(w)$ be a function that is non-decreasing, continuous, and piecewise differentiable on $[0, \bwt],$ in which
$$
\bwt = {hb \over r+h},
$$
except that $\bPhit$ might not be differentiable at $0$.  Suppose $\Phis$ satisfies the following variational inequality on $(0, \bwt]$:
$$
\la \, \bPhit = rw \, \bPhit_w +  \max \left[ \la - h (b-w) \, \bPhit_w, \, 0  \right],
\eqno(3.4)
$$
in which we use one-sided derivatives, if needed.  Additionally, suppose $\bPhit(\bwt) = 1$. Then, on $[0, \bwt]$,
$$
\bphit = \bPhit.
\eqno{\square}
$$}

So that what follows does not seem like ``mathematical magic,'' we discuss how we obtained the solution to our maximization problem.   Because we have a boundary condition at $w = \bwt$, we worked backwards from that point.  At any wealth level $w$, the individual chooses either to buy insurance of $b - w$ or to buy no insurance.  First, suppose that in a neighborhood of $\bwt$, the individual buys full insurance of $b - w$; denote the resulting solution of $\la (\bphit - 1) = - (hb - (r + h)w) \bphit_w$, with $\bphit(\bwt) = 1$, by $\bphit_f$.  Then, $\bphit_f$ is given by
$$
\bphit_f(w) = 1 - k \left( {hb - (r+h)w \over hb} \right)^{{\la \over r+h}},
$$
for $w$ near $\bwt$.  Here, $k > 0$ is some (unknown) constant.  Next, suppose that in a neighborhood of $\bwt$, the individual buys no insurance; denote the resulting solution of $\la \bphit = rw \bphit_w$, with $\bphit(\bwt) = 1$, by $\bphit_0$.  Then, $\bphit_0$ is given by
$$
\bphit_0(w) = \left( {(r+h)w \over hb} \right)^{{\la \over r}},
$$
for $w$ near $\bwt$.

To determine which of $\bphit_0$ and $\bphit_f$ is larger for $w$ near $\bwt$, compare their derivatives at $\bwt$.  Because ${\la \over r+h} < 1$, $\lim_{w \to \bwt} (\bphit_f)_w(w) = \infty$, while $(\bphit_0)_w(\bwt)$ is positive, but finite.  Thus, for wealth near $\bwt$, $\bphit_0 \ge \bphit_f$.  It might be that on some interval of wealth, we have $\bphit_0 \le \bphit_f$.  However, the existence of such an interval depends on whether $\la \le r$ or $\la > r$.  If $\la \le r$, then $\bphit_0 \ge \bphit_f$ for all $0 \le w \le \bwt$; and, in Proposition 3.2, we prove that $\bphit = \bphit_0$.  If $\la > r$, then there is a wealth level $w^* \in (0, \bwt)$ such that $\bphit_0 \le \bphit_f$ on $[0, w^*)$ and $\bphit_0 \ge \bphit_f$ on $[w^*, \bwt]$; and, in Proposition 3.5, we prove that $\bphit = \bphit_f$ on $[0, w^*)$ and $\bphit = \bphit_0$ on $[w^*, \bwt]$, with the constant $k$ chosen to make $\bphit$ continuous at $w^*$.

\prop{3.2} {If $\la \le r$, then the maximum probability of reaching the bequest goal before ruining is given by
$$
\bphit(w) = \left( {(r+h)w \over hb} \right)^{{\la \over r}},
\eqno(3.5)
$$
for initial wealth $w \in [0, \bwt]$.  The associated optimal life insurance purchasing strategy is not to purchase any life insurance until wealth reaches the safe level $\bwt$, at which time it is optimal to buy life insurance of $b - \bwt = {rb \over r+h}$.}

\pf  We use Lemma 3.1 to prove this proposition.  First, note that $\bphit$ in (3.5) is continuous and increasing on $[0, \bwt]$, $\bphit$ is differentiable on $(0, \bwt]$, and $\bphit(\bwt) = 1$.  Next, note that
$$
\la \bphit = rw \, \bphit_w,
$$
on $(0, \bwt]$.  The inequality
$$
\la - h (b-w) \, \bphit_w \le 0
$$
holds on $(0, \bwt]$ if and only if
$$
1 - {r+h \over r} \, x^{{\la \over r} - 1} + {h \over r} \, x^{{\la \over r}} \le 0,
\eqno(3.6)
$$
for all $0 < x \le 1$, in which $x = {(r+h)w \over hb}$.  For $\la = r$, inequality (3.6) is clearly true.  To show inequality (3.6) for $\la < r$, define $f$ on $(0, 1]$ by
$$
f(x) = 1 - (c+1) x^{a - 1} + c x^a,
$$
in which $c := {h \over r} > 0$ and $a := {\la \over r} \in (0, 1)$, and we wish to show that $f(x) \le 0$ on $(0, 1]$.  To this end, observe that $\lim_{x \to 0+} f(x) = -\infty$, and $f(1) = 0$, so it is enough to show that $f$ is increasing on $(0, 1]$.
$$
f'(x) = (c+1)(1-a) x^{a-2} + ca x^{a-1}
$$
is positive on $(0, 1]$ if and only if $(c+1)(1-a) \ge 0$, which is true.  Therefore, we have shown that $\bphit$ in (3.5) satisfies the variational inequality (3.4).  The optimal insurance strategy follows from the fact that $\bphit$ solves the variational inequality with $D \equiv 0$.  \qed

\rem{3.1} {When the force of mortality is less than or equal to the force of interest, the individual feels as if she has time to reach the safe level; therefore,  it is optimal for the individual to invest in the riskless asset and wait until she reaches the safe level before she buys any life insurance.  For initial wealth $w$, wealth at time $t$ equals $W(t) = w e^{rt}$, and the time that wealth reaches the safe level equals
$$
\tau_{\bwt} = {1 \over r} \ln \left( {hb \over (r+h)w} \right).
$$
The probability of reaching the safe level before dying equals $e^{-\la \tau_{\bwt}}$, which equals $\bphit$ in (3.5), as expected.  \qed}

From the discussion in Remark 3.1, we obtain the following corollary.

\cor{3.3} {If $\la \le r$, then expected wealth at death, $\bar E^t(w) = \E^{w}(W_(\tau_d))$, for an individual who follows the optimal life insurance purchasing strategy of Proposition $3.2$ is given by
$$
\bar E^t(w) =
\cases{ b \left[ 1 + {h \over r+h} \; {\la \over r - \la} \right] \left( {(r+h)w \over hb} \right)^{{\la \over r}} - {\la w \over  r - \la}, &if $\la < r$, \cr
 w \left[ {r+h \over h} +  \ln \left( {hb \over (r+h)w} \right) \right], &if $\la = r$.
} 
\eqno(3.7)
$$}

\rem{3.2} {Note that $\bar E^t$ in (3.7) uniquely solves the following BVP on $(0, \bwt]$:
$$
\left\{
\eqalign{
& \la (\bar E^t - w) =  r w \, \bar E^t_w, \cr
& \bar E^t(\bwt) = b.
}
\right.
\eqno{\square}
$$}

Next, we consider the slightly more complicated case of $\la > r$ and present a helpful lemma, whose proof we relegate to Appendix A.

\lem{3.4} {Suppose $c$ and $a$ are constants such that $0 < c < 1 < a$.  Then, the following three statements hold:
\item{$(a)$} The function $f_1$ on $[0, 1]$ defined by
$$
f_1(x) = x^a + (1-x)^c - 1
$$
has a unique zero $x^*$ in the interior $(0, 1)$.  Furthermore, $f_1(x) \le 0$ for $0 \le x \le x^*$, and $f_1(x) \ge 0$ for $x^* \le x \le 1$.
\item{$(b)$} The function $f_2$ on $[0, 1)$ defined by
$$
f_2(x) = 1 - {c \over a} \, (1 - x)^{c-1} - \left( 1 - {c \over a} \right) (1 - x)^c
$$
is non-negative on $[0, x^*]$.
\item{$(c)$} The function $f_3$ on $[0, 1]$ defined by
$$
f_3(x) = 1 - {a \over c} \, x^{a-1} + \left( {a \over c} - 1 \right) x^a
$$
is non-positive on $[x^*, 1]$.}

We use Lemma 3.4 to prove the following proposition.

\prop{3.5}  {If $\la > r$, then the maximum probability of reaching the bequest goal before ruining is given by
$$
\bphit(w) = 
\cases{
1 - \left( {hb - (r+h)w \over hb} \right)^{{\la \over r+h}}, &if $0 \le w < w^*,$ \cr
\left( {(r+h)w \over hb} \right)^{{\la \over r}}, &if $w^* \le w \le \bwt = {hb \over r+ h},$
}
\eqno(3.8)
$$
for initial wealth $w \in [0, \bwt]$.  Here, $w^*$ is the unique zero in $(0, \bwt)$ of the following expression:
$$
\left( {(r+h)w \over hb} \right)^{{\la \over r}} + \left( {hb - (r+h)w \over hb} \right)^{{\la \over r+h}} - 1.
\eqno(3.9)
$$
The associated optimal life insurance purchasing strategy is as follows:
\item{$(a)$} If wealth $w$ is less than $w^*$, then purchase life insurance of $b - w$.
\item{$(b)$} If wealth is greater than or equal to $w^*$, then do not purchase life insurance until wealth reaches the safe level $\bwt$, at which time it is optimal to buy life insurance of $b - \bwt = {rb \over r+h}$.}

\pf  First, use Lemma 3.4(a) to prove that the expression in (3.9) has a unique zero in $(0, \bwt)$.  To this end, let $a = {\la \over r} > 1$, $c = {\la \over r+h} \in (0, 1)$, and $x = {(r+h)w \over hb}$; then, the expression in (3.9) becomes $f_1$ in Lemma 3.4(a).  We know that $f_1$ has a unique zero $x^*$ in $(0, 1)$; thus, $w^* = {hb x^* \over r+h}$ is the unique zero of (3.9) in $(0, \bwt)$.

Next, note that $\bphit$ is non-decreasing, continuous, and piecewise differentiable on $[0, \bwt]$, with $\bphit(\bwt) = 1$.  On $[0, w^*)$,
$$
(\la - 1) \bphit = ((r+h)w - b) \bphit_w,
\eqno(3.10)
$$
and the inequality
$$
\la - h(b-w) \bphit_w \ge 0
\eqno(3.11)
$$
holds if and only if
$$
1 - {b - w \over b} \left( 1 - {(r+h)w \over hb} \right)^{{\la \over r+h} - 1} \ge 0.
\eqno(3.12)
$$
In inequality (3.12), let $a = {\la \over r} > 1$, $c = {\la \over r+h} \in (0, 1)$, and $x = {(r+h)w \over hb}$, as before; then, (3.12) becomes $f_2(x) \ge 0$ on $[0, x^*)$, which we know is true from Lemma 3.4(b).  Thus, we have proved inequality (3.11) on $[0, w^*)$.  From equation (3.10) and inequality (3.11), it follows that $\bphit$ satisfies the variational inequality (3.4) in Lemma 3.1 on $[0, w^*)$.  Because $\bphit$ satisfies (3.3) with $D(w) = b-w$ when $0 \le w < w^*$, we deduce that for wealth less than $w^*$, it is optimal to buy insurance in order to reach the bequest goal $b$.

On $[w^*, \bwt]$,
$$
\la \bphit = rw \, \bphit_w,
\eqno(3.13)
$$
and the inequality
$$
\la - h (b-w) \, \bphit_w \le 0
\eqno(3.14)
$$
holds if and only if
$$
1 - {r+h \over r} \cdot {b - w \over b} \left( {(r+h)w \over hb} \right)^{{\la \over r} - 1} \le 0.
\eqno(3.15)
$$
In inequality (3.15), let $a = {\la \over r} > 1$, $c = {\la \over r+h} \in (0, 1)$, and $x = {(r+h)w \over hb}$, as before; then, (3.15) becomes $f_3(x) \le 0$ on $[x^*, 1]$, which we know is true from Lemma 3.4(c).  Thus, we have proved inequality (3.14) on $[w^*, \bwt]$.  From equation (3.13) and inequality (3.14), it follows that $\bphit$ satisfies the variational inequality (3.4) in Lemma 3.1 on $[w^*, \bwt]$.  Because $\bphit$ satisfies (3.3) with $D(w) = 0$ when $w^* \le w \le 1$, we deduce that for wealth greater than or equal to $w^*$, it is optimal not to buy insurance.  Instead, it is optimal to wait until wealth reaches the safe level $\bwt = {hb \over r+h}$, at which time the individual will buy insurance of ${rb \over r+h}$.  \qed

\rem{3.3} {For wealth equal to $w^*$, the probability that wealth at death equals $b$ is the same whether the individual buys full insurance $D(w) = b - w$ until wealth reaches 0 or whether she buys no insurance until wealth reaches the safe level.  So, she is indifferent between these two strategies, and we picked the buy-no-insurance strategy because her expected wealth at death is greater under that strategy; see Corollary 3.6 below.}

\rem{3.4} {When $\la > r$ and when initial wealth $w \in [0, w^*)$, optimally controlled wealth at time $t$ equals
$$
W(t) = {hb \over r+ h} - \left( {hb \over r+ h} - w \right) e^{(r+h)t},
$$
which continually decreases and might reach zero before the individual dies.  The time that wealth hits zero is given by
$$
\tau_0 = {1 \over r+h} \ln \left( {hb \over hb - (r+h) w} \right).
$$
The probability that the individual dies with wealth at death (including death benefit) equal to $b$ equals the probability that the individual dies before time $\tau_0$, or $1 - e^{-\la \tau_0}$, which equals $\bphit$, as expected.  When initial wealth $w \in [w^*, \bwt]$, then the individual invests all her wealth in the riskless asset, so that wealth at time $t$ equals $w e^{rt}$, and she does not buy insurance until wealth reaches the safe level $\bwt = {hb \over r+h}$.  \qed}

From the discussion in Remark 3.4, we obtain the following corollary.

\cor{3.6} {If $\la > r$, then expected wealth at death, $\bar E^t(w) = \E^{w}(W(\tau_d))$, for an individual who follows the optimal life insurance purchasing strategy of Proposition $3.4$ is given by
$$
\bar E^t(w) =
\cases{
b \left[ 1 - \left( {hb - (r+h)w \over hb} \right)^{{\la \over r+h}} \right], &if $0 \le w < w^*,$ \cr
b \left[ 1 - {h \over r+h} \; {\la \over \la - r} \right] \left( {(r+h)w \over hb} \right)^{{\la \over r}} + {\la w \over  \la - r}, &if $w^* \le w \le \bwt.$
}
\eqno(3.16)
$$
Moreover, $\bar E^t(w^*-) < \bar E^t(w^*+)$.}



\rem{3.5} {On $[0, w^*)$, one can show that $\bar E^t$ in (3.16) uniquely solves the following BVP:
$$
\left\{
\eqalign{
& \la (\bar E^t - b) =  ((r + h) w - hb) \, \bar E^t_w, \cr
& \bar E^t(0) = 0.
}
\right.
$$
On $[w^*, \bwt]$,  $\bar E^t$ in (3.16) uniquely solves the following BVP, as in Remark 3.2:
$$
\left\{
\eqalign{
& \la (\bar E^t - w) =  r w \, \bar E^t_w, \cr
& \bar E^t(\bwt) = b.
}
\right.
\eqno{\square}
$$}

Next, we present properties of the dividing point $w^*$.  In the interest of space, we omit the proof of this corollary but  invite the interested reader to provide it.

\cor{3.7} {When $\la > r$, the dividing point $w^*$ between full insurance $D(w) = b - w$ for $w < w^*$ and no insurance $D(w) = 0$ for $w \ge w^*$ satisfies the following properties: 
\smallskip
\item{$(a)$} $w^*$ increases proportionally with respect to $b,$ the bequest goal.
\item{$(b)$} $w^*$ increases with respect to $\la,$ the force of mortality.
\item{$(c)$} $w^*$ decreases with respect to $r,$ the riskless rate of return.}

\rem{3.6} {\item{(a)} It is clear that $w^*$ changes proportionally with respect to $b$.  Indeed, $x^* = {(r+h)w^* \over hb}$ solves \break $1 - (x^*)^{{\la \over r}} = (1-x^*)^{{\la \over r+h}}$.  This equation is independent of $b$, so $x^*$ does not change with $b$; thus, ${w^* \over b}$ is constant.
\item{(b)} There are competing effects of $\la$ on $w^*$.  For the premium rate $h$ fixed, $w^*$  increases with $\la$ because the individual is more likely to die before reaching the safe level $\bwt = {hb \over r+h}$.  Thus, for $h$ fixed, she is more likely to want to buy full insurance now instead of waiting to reach the safe level.  However, the premium rate $h$ increases with $\la$, so we have to consider how $w^*$ changes with $h$.  The safe level increases with $h$, which makes the individual less willing to wait to reach the safe level.  But, the premium becomes more expensive as $h$ increases; thus, the individual's desire to buy full insurance is dampened.   The net of these effects is that $w^*$ increases with $\la$; that is, the extra cost of the premium is not enough to fully eliminate the individual's greater willingness to buy full insurance now.
\item{(c)} There are two re-enforcing effects of $r$ on $w^*$.  First, the safe level ${hb \over r+ h}$ decreases with $r$, so the individual does not have to wait as long to reach the safe level.  Second, if $r$ increases, then the individual's money increases at a faster rate (namely, $r$) and reaches any level sooner.  Thus, $w^*$ decreases with $r$ because the individual is more willing to wait to reach the safe level.}
\hangindent 20 pt \item{(d)}  We found examples that demonstrate that $w^*$ might decrease with $\bt$ or might increase with $\bt$.  There are two competing effects of $\bt$ on $w^*$ as discussed in part (b) above: increasing the safe level versus increasing the premium.  If the effect of increasing the premium is larger than that of increasing the safe level, then $w^*$ decreases with $\bt$, and vice versa.

\ex{3.1} {To demonstrate the discussion in Remark 3.6, we present this numerical example.  For our base scenario, take the following parameter values:  $b = 1$, $r = 0.03$, $\la = 0.08$, and $h = (1+0.25)\la = 0.10$.  Thus, the safe level equals ${10 \over 13} = 0.7692$, and $w^* = 0.6949$.
\item{(a)} First, we show numerically that $w^*$ increases with respect to $\la$, while keeping $\bt = 0.25$.  Write $w^*(\la)$ to denote $w^*$'s dependence on $\la$ in this example.  We have $w^*(0.04) = 0.0873$, $w^*(0.05) = 0.3323$, $w^*(0.06) = 
0.5118$, and $w^*(0.08) = 0.6949$.
\item{(b)} Next, we show numerically that $w^*$ decreases with respect to $r$.  Write $w^*(r)$ to denote $w^*$'s dependence on $r$ in this example.  Recall that $r < \la$, so we will consider $r \in [0, 0.08)$.  We have $w^*(0.00) = 1.0000$, $w^*(0.01) = 0.9091$, $w^*(0.02) = 0.8333$, $w^*(0.03) = 0.6949$, $w^*(0.04) = 0.5118$, $w^*(0.05) = 0.2864$, $w^*(0.06) = 0.0873$, and $w^*(0.07) = 0.0030$.}
\hangindent 20 pt \item{(c)} Finally, we show numerically that $w^*$ might decrease or increase with respect to $h$, while keeping $\la$ fixed--that is, we vary $\bt$. Write $w^*(h)$ to denote $w^*$'s dependence on $h$ in this example.  First, $w^*(0.08) = 0.7054$, $w^*(0.10) = 0.6949$, and $w^*(0.15) = 0.6197$.  So, $w^*$ decreases with $h$ when $\la = 0.08$.  Next, let $\la = 0.12$; then, $w^*(0.12) = 0.7992$, $w^*(0.15) = 0.8193$, $w^*(0.20) = 0.8101$, and $w^*(0.25) = 0.7838$.  So, $w^*$ first increases and then decreases with $h$ when $\la = 0.12$.

\rem{3.7} {We now summarize what we have learned in this section and clarify some of the results.  Suppose the individual decides to start buying full insurance at a wealth level $w$ that is less than the safe level. This is a  winning move if she dies  before time $\tau_0$, the time at which her wealth is depleted to zero; on the other hand, waiting to buy until after reaching the safe level is the winning move if she lives to time $\tau_{\bwt}$. Therefore, by letting $p(t) = e^{-\la t}$ denote the probability of living to time $t$, the better strategy is to buy full insurance if $1-p(\tau_0) \ge  p(\tau_{\bwt})$, that is,  $p(\tau_0) + p(\tau_{\bwt}) \le 1$, while the better strategy is to wait if the inequality goes the other way.
        
We see, therefore, that $w^*$ is precisely the wealth level that results  in
$$
p(\tau_0) + p(\tau_{\bwt}) = 1.
$$
We  can see from this equation that any changes that cause both $\tau_0$ and $\tau _{\bwt}$ to increase will decrease both probabilities and thereby increase $w^*$, while changes that cause both times to decrease (such as an increase in $r$) will decrease $w^*$.  For changes that cause the two times to move in different directions, the effect can be uncertain, as we noticed above for an increase in $h$, which causes $\tau_{\bwt}$ to increase but $\tau_0$ to decrease.}

\subsect{3.2  Irreversible whole life}

In this section, we assume that once the individual buys a given amount of insurance $D$, then she must pay premium at the rate of $hD$ for the remainder of her life.  She cannot reverse this purchase.  Wealth follows the process given in (3.1).  Denote the maximum probability of dying with wealth at least $b$ before ruining by $\bphi$; it is defined as in (3.2), except that the definition of admissible strategy differs in this case.   Indeed, an insurance purchasing strategy $\D = \{ D(t) \}_{t \ge 0}$ is {\it admissible} if $\D$ is a non-negative and non-decreasing process. 

In the case for irreversible whole life insurance with premium payable continuously, the safe level differs depending on the existing amount of life insurance $D$.  Indeed, for a given level of wealth $w$, the individual can safely invest it in the riskless asset and earn investment income at the rate of $rw$. Because the individual already has a death benefit of $D$, at the safe level, this income must be sufficient to cover the insurance premium; that is, $rw \ge hD$, or equivalently, $w \ge {hD \over r}$.

Moreover, if $D$ is less than ${rb \over r+h}$, then we have the safe level from Section 3.1, namely ${hb \over r+h}$.  Thus, the safe level when life insurance is irreversible is given by
$$
\bw(D) = \max \left[ {hb \over r + h}, {hD \over r} \right] =
\cases{{hb \over r + h}, &if $D \le {rb \over r + h}$, \cr
{hD \over r}, &if $D > {rb \over r + h}$.}
$$
The verification lemma for $\bphi$ is as follows.

\lem{3.8} {Let $\bPhi = \bPhi(w, D)$ be a function that is non-decreasing, continuous, and piecewise differentiable with respect to both $w$ and $D$ on $\bDo = \{ (w, D): 0 \le w \le \bw(D), \,  D \ge 0 \}$.  Suppose $\bPhi$ satisfies the following variational inequality on $\bDo$:
$$
\max \left( (rw - hD) \bPhi_w - \la \left( \bPhi - {\bf 1}_{\{w+D \ge b\}} \right), \, \bPhi_D \right) = 0,
\eqno(3.17)
$$
in which we use one-sided derivatives, if needed. Additionally, suppose $\bPhi(\bw(D), D) = 1$. Then, on $\bDo$,
$$
\bphi = \bPhi.
\eqno{\square}
$$}

When $D \ge b$, the optimal life insurance purchasing strategy is not to buy any additional life insurance because $D$ already meets the targeted bequest.  The goal for the individual is not to ruin while paying the premium rate $hD$.  Thus, in this case, $\bphi$ solves the following BVP:
$$
\left\{
\eqalign{
& \la \left( \bphi - 1 \right) = (rw - hD) \bphi_w, \cr
& \bphi \left({hD \over r}, D \right) = 1.}
\right.
$$
We give the solution to this BVP in the next proposition; we leave it to the reader to show that it satisfies variational inequality (3.17) of Lemma 3.8.

\prop{3.9} {On $\bDo_0 = \{ (w, D): 0 \le w \le {hD \over r}, \, D \ge b \}$, the maximum probability of reaching the bequest goal before ruining is given by
$$
\bphi(w, D) = 1 - \left( {hD - rw \over hD} \right)^{{\la \over r}}.
\eqno(3.18)
$$
The associated optimal insurance purchasing strategy is not to buy any additional insurance. \qed}


\rem{3.8} {When $D \ge b$ and when initial wealth $w$ lies in $\left[0, {hD \over r} \right]$, optimally controlled wealth at time $t$ equals
$$
W(t) = {hD \over r} - \left( {hD \over r} - w \right) e^{rt},
$$
which decreases over time.  Thus, wealth will never reach the safe level, and the relevant hitting time is the hitting time of zero wealth, $\tau_0$, which equals
$$
\tau_0 = {1 \over r} \ln \left( {hD \over hD - rw} \right).
$$
The probability that the individual dies with wealth at death equal to at least $b$ is the probability that she dies before time $\tau_0$, or $1 - e^{-\la \tau_0}$, which equals (3.18), as expected.

There is an interesting analogy between the case for which $D \ge b$ and the case, in Section 3.1, for which $\la > r$ and initial wealth $w \in [0, w^*)$.  Indeed, by examining the above expression for $W(t)$ with the one given in Remark 3.4, we see that we can get the former from the latter by replacing $D$ and $r$ with $b$ and $r + h$, respectively.  The hitting times of zero similarly correspond, as do the probabilities of dying before wealth reaches 0.  In other words, we can get (3.18) from the first expression in (3.8) by replacing $b$ and $r+h$ with $D$ and $r$, respectively.  \qed}

From the discussion in Remark 3.8, we obtain the following corollary.

\cor{3.10} {If $D \ge b$, then expected wealth at death, $\bE(w, D) = \E^{w, D} (W(\tau_d))$, for an individual who follows the optimal life insurance purchasing strategy of Proposition $3.9$ is given by
$$
\bE(w, D) = \cases{
D \left( 1 - {h \over \la - r} \right) \left( 1 - \left( { hD - rw \over hD} \right)^{{\la \over r}} \right) + {\la w \over \la - r}, &if $\la \ne r$, \cr \cr
{hD - rw \over r} \ln \left( { hD - rw \over hD} \right) + {(r+h)w \over h}, &if $\la = r$.
}
\eqno(3.19)
$$}

\rem{3.9} {If $D \ge b$, then $\bE$ in (3.19) uniquely solves the following BVP for $0 \le w \le {hD \over r}$:
$$
\left\{
\eqalign{
& \la (\bE - (w + D)) =  (rw - hD) \, \bE_w, \cr
& \bE \left( 0, D \right) = 0.
}
\right.
\eqno{\square}
$$}

Henceforth, we assume that $D < b$, and we will require that $\bphi$ be continuous across $D = b$.  We propose the following ansatz for optimally purchasing life insurance, in which $w$ and $D$ are initial wealth and death benefit, respectively:

\smallskip

\item{(a)}  Suppose $b - D \le w \le {hD \over r}$ and ${rb \over r+h} < D < b$; then, hypothesize that the individual will buy no additional life insurance if $w > b - D$.  If wealth reaches the value $b - D$, then, via instantaneous control of the death benefit, wealth and death benefit will stay on the line $w' + D' = b$, moving toward the point $(w', D') = (0, b)$.

\item{(b)} Suppose $0 \le D < b-w$ and $0 \le w \le {hb \over r+h}$.

\itemitem{(i)}  Hypothesize that if $(w, D)$ is ``close enough'' to the line $w' + D' = b$, then the individual will buy additional life insurance of $b - (w + D)$ and thereafter will keep wealth and death benefit on the line $w' + D' = b$.  We expect points on the line $rw' = hD'$ to lie in this ``jump'' region; otherwise, from the differential equation in (3.17), we have $\bphi = 0$ along $rw' = hD'$, which is not true.

\itemitem{(ii)} Hypothesize that if $w$ is ``close enough'' to the safe level ${hb \over r+h}$, then the individual will buy no additional insurance until her wealth reaches the safe level.  Inherent in this part of the ansatz is that $D < {rb \over r+h}$, so that the safe level equals $w = {hb \over r+h}$.

\bigskip

We will (slightly) abuse notation below by referring to $\bphi$ as the solution of various boundary-value problems resulting from the above ansatz.  However, as we progress, we will prove that the $\bphi$ we thus obtain is indeed the maximum probability of reaching the bequest goal before ruining.

\medskip

\noi {\bf Region }  $\bDo_a = \{ (w, D): b - D \le w \le {hD \over r}, \; {rb \over r+h} < D < b\}$:  Based on part (a) of the ansatz, the maximum probability of reaching the bequest before ruining solves the following BVP:
$$
\left\{
\eqalign{ &\la( \bphi - 1) = (rw - hD) \bphi_w, \cr
& \bphi_D(b-D, D) = 0, \cr
& \bphi \left( {hD \over r}, D \right) = 1, \qquad \lim_{D \to b-} \bphi(w, D) = 1 - \left( {hb - rw \over hb} \right)^{{\la \over r}}.
}
\right.
\eqno(3.20)
$$
The condition that $\bphi_D = 0$ at $w = b - D$ arises from the ansatz that the individual purchases insurance continuously along that line.  The last condition comes from requiring continuity at $D = b$; alternatively, we could simply require that $\lim_{(w,D) \to (0+, b-)} \bphi(w, D) = 0$ and then check that continuity at $D = b$ holds.  The solution of (3.20) is given by
$$
\bphi(w, D) = 1 - \left( {(r+h)D - rb \over hb} \right)^{{\la \over r+h}}  \left( {hD - rw \over (r+h)D - rb} \right)^{{\la \over r}}.
\eqno(3.21)
$$

In the next proposition, we state that $\bphi$ in (3.21) equals the maximum probability of reaching the bequest goal; one can prove this proposition via Lemma 3.8.

\prop{3.11} {On $\bDo_a = \{ (w, D): b - D \le w \le {hD \over r}, \, {rb \over r+h} < D < b \}$, the maximum probability of reaching the bequest goal before ruining is given by $\bphi$ in $(3.21)$. The associated optimal insurance purchasing strategy is to buy additional insurance only when wealth reaches $b - D$, after which continually buy additional insurance to ensure that the sum of wealth and death benefit equals $b$.}


\rem{3.10} {For initial wealth and death benefit lying in the interior of $\bDo_a$, optimally controlled wealth at time $t$ equals
$$
W(t) = {hD \over r} - \left( {hD \over r} - w \right) e^{rt},
$$
which decreases over time.  Thus, wealth will never reach the safe level, and the first relevant hitting time is the time that wealth reaches $b - D$, $\tau_{b-D}$, which equals
$$
\tau_{b-D} = {1 \over r} \ln \left( {(r+h)D - rb \over hD - rw} \right).
$$

After wealth reaches $b - D$, the individual continually buys life insurance to keep wealth plus death benefit equal to $b$.  It follows from Remark 3.4, that at time $t = \tau_{b-D} + s$ for $s \ge 0$, optimally controlled wealth equals
$$
W(t) = {hb \over r+ h} - \left( D - {rb \over r+ h} \right) e^{(r+h)s},
$$
which decreases over time.  Thus, the second relevant hitting time is the hitting time of zero,
$$
\tau_0 = {1 \over r+h} \ln \left( {hb \over (r+h)D - rb} \right),
$$
which we measure from time $\tau_{b-D}$.  The probability that the individual dies with wealth at death of at least $b$ equals the probability that she dies before time $\tau_{b-D}$ plus the probability that she dies before time $\tau_0$ given that she dies after time $\tau_{b-D}$, or $\left( 1 - e^{-\la \tau_{b-D}} \right) + e^{-\la \tau_{b-D}} \left(1 - e^{-\la \tau_0} \right) = 1 - e^{-\la (\tau_0 + \tau_{b-D})}$, which equals the expression in (3.21), as expected.  \qed}

From the discussion in Remark 3.10, we obtain the following corollary.

\cor{3.12} {For $(w, D) \in \bDo_a$, expected wealth at death for an individual who follows the optimal life insurance purchasing strategy of Proposition $3.11$ is given by
$$
\bE(w, D) = 
\cases{ \left( {hD - rw \over (r+h)D - rb} \right)^{{\la \over r}} \left[ {(r+h)D \over \la - r}  - b \left\{ {r \over \la - r} + \left( {(r+h)D - rb \over hb} \right)^{{\la \over r+h}} \right\} \right] + D \left[ 1 - {h \over \la - r} \right] + {\la w \over \la - r}, &if $\la \ne r$, \cr
w + D -  {hD - rw \over r} \, \ln \left( {(r+h)D - rb \over hD - rw} \right) - {hD - rw \over h} \left( {hb \over (r+h)D - rb} \right)^{h \over r+h}, &if $\la = r$.}
\eqno(3.22)
$$}


\rem{3.11} {For $(w, D) \in \bDo_a$, $\bE$ in (3.22) uniquely solves the following BVP:
$$
\left\{
\eqalign{
& \la (\bE - (w + D)) =  (rw - hD) \, \bE_w, \cr
& \bE_D \left( b-D, D \right) = 0, \qquad \bE(w, b) = \bar e(w),
}
\right.
$$
in which
$$
\bar e(w) = 
\cases{b \left( 1 - {h \over \la - r} \right) \left( 1 - \left( { hb - rw \over hb} \right)^{{\la \over r}} \right) + {\la w \over \la - r}, &if $\la \ne r$, \cr \cr
{hb - rw \over r} \ln \left( { hb - rw \over hb} \right) + {(r+h)w \over h}, &if $\la = r$.}
$$
The boundary condition at $D = b$ comes from continuity of $\bE$ across $D = b$; thus, $\bar e$ is obtained via the expressions in (2.22a) with $D = b$.  Alternatively, we could simply require that $\lim_{(w, D) \to (0+, b-)} \bE(w, D) = 0$ and then check that continuity at $D = b$ holds.   \qed}

\medskip

\noi {\bf Region } $\bDo_b = \{ (w, D): 0 \le D < b - w, \; 0 \le w \le {hb \over r+h}\}$:  Based on part (b)(i) of the ansatz, for $(w, D)$ ``close enough'' to the line $w' + D' = b$, the individual immediately buys additional life insurance of $b - (w+D)$.   Thus,  $\bphi$ is given by $\bphi(w, D) = \bphi(w, b-w)$, in which the right side is given by (3.21).  Thus,
$$
\bphi(w, D) = 1 -  \left( {hb - (r+h)w \over hb} \right)^{{\la \over r+h}}.
\eqno(3.23)
$$

Based on part (b)(ii) of the ansatz, for $w$ ``close enough'' to the safe level ${hb \over r+h}$, assuming that $D < {rb \over r+h}$, $\bphi$ solves the following BVP:
$$
\left\{
\eqalign{ &\la \bphi = (rw - hD) \bphi_w, \cr
& \bphi \left( {hb \over r+h}, D \right) = 1.
}
\right.
\eqno(3.24)
$$
The solution of (3.24) is given by
$$
\bphi(w, D) = \left( {rw - hD \over h\left( {rb \over r+ h} - D \right)} \right)^{{\la \over r}}.
\eqno(3.25)
$$
To obtain (3.25), we assume that $rw - hD > 0$ when $w < {hb \over r+h}$; otherwise, if the line $rw = hD$ is in the continuation region, the differential equation in (3.24) implies that $\bphi = 0$ along $rw = hD$, which is not true.  Also, in writing (3.25), we mean that $\bphi = 1$ if $w = {hb \over r+h}$ and $D = {rb \over r+h}$ because that point is in the safe region.

Next, we find the boundary between the jump region underlying the expression in (3.23) and the continuation region underlying the expression in (3.25).  It turns out that we can express this boundary as a function $D = D_j(w)$; subscript {\it j} for jump.  We require $\bphi$ to be continuous along that boundary; that is, we require
$$
1 -  \left( {hb - (r+h)w \over hb} \right)^{{\la \over r+h}} = \left( {rw - hD_j(w) \over h\left( {rb \over r+ h} - D_j(w) \right)} \right)^{{\la \over r}}.
$$
Solving this equation for $D_j$ for $0 \le w < {hb \over r+h}$ yields
$$
D_j(w) = {r \over h} \, {w - {hb \over r+ h} f_j(w) \over 1 - f_j(w)},
\eqno(3.26)
$$
in which $f_j$ is given by
$$
f_j(w) = \left[ 1 - \left( {hb - (r+h)w \over hb}  \right)^{{\la \over r+h}} \right]^{{r \over \la}}.
\eqno(3.27)
$$
Because $f_j \left( {hb \over r+h} \right) = 1$, we define $D_j \left( {hb \over r+h} \right)$ by continuity; specifically, set $D_j \left( {hb \over r+h} \right) = {rb \over r+h}$.  For what comes later, it is important to understand the graph of $D_j$ on $\left[0, {hb \over r+h} \right]$.  See Appendix B for the proof of the following lemma.

\lem{3.13} {Let the function $D = D_j(w)$ be defined by equations $(3.26)$ and $(3.27),$ for $0 \le w < {hb \over r+h},$ and define $D_j \left( {hb \over r+h} \right) = {rb \over r+h}$.
\item{$(a)$}  $D_j(w) \le {rw \over h}$, with equality only at $w = 0$ and $w = {hb \over r+h}$.
\item{$(b)$}  If $\la \le r$, then $D_j(w)$ increases from $0$ to ${rb \over r+h}$ as $w$ increases from $0$ to ${hb \over r+h}$.
\item{$(c)$}  If $\la > r$, then $D_j(w) \le 0$ for $0 \le w \le w^*$, and $D_j(w)$ increases from $0$ to ${rb \over r+h}$ as $w$ increases from $w^*$ to ${hb \over r+h}$, in which $w^*$ is the unique zero of the expression in $(3.9)$ in Proposition $3.4$.}

From Lemma 3.13, we see that there are two cases to consider: $\la \le r$ and $\la > r$, as there were for the problem in Section 3.1.  In the next two propositions, we prove that $\bphi$ given in (3.23) and (3.25), patched together along $D = D_j(w)$, satisfies the conditions of Lemma 3.8.  In the first, we consider $\la \le r$; in the second, $\la > r$.

\prop{3.14} {Suppose $\la \le r$.  On $\bDo_b = \{ (w, D): 0 \le D < b - w, \; 0 \le w \le {hb \over r+h} \}$,  the maximum probability of reaching the bequest goal before ruining is given by
$$
\bphi(w, D) = 
\cases{\left( {rw - hD \over h\left( {rb \over r+ h} - D \right)} \right)^{{\la \over r}}, &if $0 \le D \le D_j(w)$, \cr
1 -  \left( {hb - (r+h)w \over hb} \right)^{{\la \over r+h}}, &if $D_j(w) < D \le b-w$.}
\eqno(3.28)
$$
The associated optimal insurance purchasing strategy is as follows:
\item{$(a)$} If $0 \le D \le D_j(w)$, then do not buy additional insurance until wealth reaches the safe level ${hb \over r+h}$, at which time, buy additional insurance of ${rb \over r+h} - D$.
\item{$(b)$} If $D_j(w) < D \le b-w$, then immediately buy additional insurance of $b - (w+D)$ and thereafter continually buy additional insurance to ensure that the sum of wealth and death benefit equals $b$.}

\pf  The function $\bphi$ is increasing and piecewise differentiable in $w$ and $D$, and it equals 1 at $w = {hb \over r+h}$.  Recall that we defined $\bphi$ in (3.25) to be equal to 1 when $w = {hb \over r+h}$ and $D = {rb \over r+h}$.   From the definition of $D_j(w)$, we know that $\bphi$ in (3.28) is continuous in $\bDo_b$;  it is also continuous with $\bphi$ given in (3.21) across the line $w + D = b$.  It remains to show that $\bphi$ satisfies the variational inequality (3.17).

If $0 \le D \le D_j(w)$, then $(rw - hD) \bphi_w - \la \bphi = 0$ by derivation of the expression in (3.25).  The inequality $\bphi_D \le 0$ holds on this subregion because, in the interior,
$$
\bphi_D(w, D) \propto {1 \over {rb \over r+h} - D} - {h \over rw-hD},
$$
which is negative because $w < {hb \over r+h}$, and because $rw > hD$ and $D < {rb \over r+h}$ when $D < D_j(w)$.

If $D_j(w) < D \le b-w$, then clearly $\bphi_D = 0$.  The inequality $(rw - hD) \bphi_w - \la \bphi \le 0$ holds on this subregion if and only if
$$
D \ge (b - w) - b \left( {hb - (r+h)w \over hb} \right)^{1 - {\la \over r+h}} =: D_0(w).
\eqno(3.29)
$$
Thus, it is enough to show that $D_0(w) \le D_j(w)$.  This inequality holds with equality at $w = 0$ and $w = {hb \over r+h}$, so we only need to show it for $0 < w < {hb \over r+h}$.  By setting $z = \left( {hb - (r+h)w \over hb} \right)^{{\la \over r+h}}$, the inequality $D_0(w) \le D_j(w)$ for $0 < w < {hb \over r+h}$  becomes
$$
(1 - z)^{1 - {r \over \la}} - 1 + {h \over r+h} \, z \ge 0,
$$
for $0 < z < 1$.  This inequality holds because the left side equals 0 for $z = 0$, and the left side increases on $(0, 1)$ because $\la \le r$.

We have, thus, shown that $\bphi$ in (3.28) satisfies the conditions of Lemma 3.8, so we conclude that $\bphi$ is the maximum probability of reaching the bequest goal before ruining.  \qed

\rem{3.12} {If $\la \le r$, then on the subregion of $\bDo_b$ for which $0 \le D \le D_j(w)$, optimally controlled wealth at time $t$ equals
$$
W(t) = {hD \over r} + {rw - hD \over r} \, e^{rt},
$$
which increases towards the safe level ${hb \over r+h}$.  The hitting time of ${hb \over r+h}$ equals
$$
\tau_{{hb \over r+h}} = {1 \over r} \, \ln \left( {h \left( {rb \over r+h} - D \right) \over rw - hD} \right).
$$
The probability that the individual dies with wealth at death equal to $b$ is the probability that she dies after time $\tau_{{hb \over r+h}}$, or $e^{-\la \tau_{{hb \over r+h}}}$, which equals the expression in (3.25), as expected.

On the subregion of $\bDo_b$ for which $D_j(w) < D \le b - w$, the individual immediately buys additional insurance of $b - (w + D)$ and thereafter continually buys insurance to stay on the line $w' + D' = b$.  From Remark 3.4, it follows that optimally controlled wealth at time $t$ equals
$$
W(t) = {hb \over r+ h} - \left( {hb \over r+ h} - w \right) e^{(r+h)t},
$$
which continually decreases and might reach zero before the individual dies.  The time that wealth hits zero depends on $w$:
$$
\tau_0 = {1 \over r+h} \ln \left( {hb \over hb - (r+h) w} \right).
$$
The probability that the individual dies with wealth at death equal to $b$ equals the probability that the individual dies before time $\tau_0$, or $1 - e^{-\la \tau_0}$, which equals the expression in (3.23), as expected.   \qed}

\cor{3.15} {If $\la \le r$, then expected wealth at death on the subregion of $\bDo_b$ for which $0 \le D \le D_j(w)$ is given by the following expression when the individual follows the optimal life insurance purchasing strategy in Proposition $3.14$:
$$
\bE(w, D) = 
\cases{
\left\{ (b-D) \left[ 1 + {h \over r+h}  {\la \over r- \la} \right] - {hD \over r - \la} \left[ 1 - {\la \over r+h} \right] \right\} \left( {rw - hD \over h \left( {rb \over r+h} - D \right) }  \right)^{{\la \over r}} + D \, { r + h - \la \over r - \la} - {\la w \over r- \la}, &if $\la < r$, \cr \cr
{rw - hD \over r} \ln \left( {h \left( {rb \over r+h} - D \right) \over rw - hD} \right) + {r+h \over r} \, w, &if $\la = r$.}
\eqno(3.30)
$$
If $\la \le r$, then expected wealth at death on the subregion of $\bDo_b$ for which $D_j(w) < D \le b - w$ is given by
$$
\bE(w, D) = b \left[ 1 - \left( {hb - (r+h)w \over hb} \right)^{{\la \over r+h}} \right].
\eqno(3.31)
$$}

\pf From the discussion in Remark 3.12, we deduce that, for $0 \le D \le D_j(w)$,
$$
\bE(w, D) = \int_0^{\tau} \left( {hD \over r} + {rw - hD \over r} \, e^{rt} + D \right) \la e^{-\la t} \, dt + b \, \bphi(w, D),
$$
from which the expressions in (3.30) follow.  To obtain (3.31) for $D_j(w) < D < b-w$, recall that $(w, D)$ immediately jumps to $(w, b - w)$ and thereafter stays on the line $w' + D' = b$.  Thus, we can use the work in Corollary 3.6 to deduce that $\bE(w, D) = \bar E^t(w)$, in which $\bar E^t$ is given by the first expression in (3.16).  \qed

\rem{3.13}  {If $\la \le r$, then expected wealth at death on the subregion of $\bDo_b$ for which $0 \le D \le D_j(w)$ solves the following BVP:
$$
\left\{
\eqalign{
& \la (\bE - (w + D)) =  (rw - hD) \, \bE_w, \cr
& \bE \left( {hb \over r+h}, D \right) = b.
}
\right.
$$
On the subregion for which $D_j(w) < D < b - w$, expected wealth at death solves the first BVP given in Remark 3.5.  \qed}

\prop{3.16} {Suppose $\la > r$.  On $\bDo_b = \{ (w, D): 0 \le D < b - w, \; 0 \le w \le {hb \over r+h} \}$,  the maximum probability of reaching the bequest goal before ruining is given by
$$
\bphi(w, D) = 
\cases{\left( {rw - hD \over h\left( {rb \over r+ h} - D \right)} \right)^{{\la \over r}}, &if $0 \le D \le D_j(w)$ and $w^* \le w \le {hb \over r+h}$, \cr
1 -  \left( {hb - (r+h)w \over hb} \right)^{{\la \over r+h}}, &otherwise.}
\eqno(3.32)
$$
Here, $w^*$ is the unique zero in $\left(0, {hb \over r+h} \right)$ of the expression in $(3.9).$  The associated optimal insurance purchasing strategy is as follows:
\item{$(a)$} If $0 \le D \le D_j(w)$, then do not buy additional insurance until wealth reaches the safe level ${hb \over r+h}$, at which time, buy additional insurance of ${rb \over r+h} - D$.
\item{$(b)$} Otherwise, immediately buy additional insurance of $b - (w+D)$ and thereafter continually buy additional insurance to ensure that the sum of wealth and death benefit equals $b$.}

\pf  The function $\bphi$ is increasing and piecewise differentiable in $w$ and $D$, and it equals 1 at $w = {hb \over r+h}$.  Recall that we defined $\bphi$ in (3.25) to be equal to 1 when $w = {hb \over r+h}$ and $D = {rb \over r+h}$.   From the definition of $D_j(w)$, we know that $\bphi$ in (3.32) is continuous in $\bDo_b$;  it is also continuous with $\bphi$ given in (3.21) across the line $w + D = b$.  It remains to show that $\bphi$ satisfies the variational inequality (3.17).

If $0 \le D \le D_j(w)$ and $w^* \le w \le {hb \over r+h}$, then $(rw - hD) \bphi_w - \la \bphi = 0$ by derivation of the expression in (3.25).  The proof of Proposition 3.14 shows us that the inequality $\bphi_D \le 0$ holds on this subregion.

If $0 \le D \le b - w$ and $0 \le w < w^*$ or if $D_j(w) < D \le b - w$ and $w^* \le w \le {hb \over r+h}$, then clearly $\bphi_D = 0$.  The inequality $(rw - hD) \bphi_w - \la \bphi \le 0$ holds on this subregion if and only if $D \ge D_0(w)$, in which $D_0$ is given in (3.29).  Thus, we must show that $D_0(w) \le 0$ for $0 \le w < w^*$ and that $D_0(w) \le D_j(w)$ for $w^* \le w \le {hb \over r+h}$.  The inequality $D_0(w) \le 0$ for $0 \le w < w^*$ is equivalent to $f_2(x) \ge 0$ for $0 \le x \le x^*$, in which $x = {(r+h)w \over hb}$ and $f_2$ and $x^*$ are as in Lemma 3.4.  From Lemma 3.4, we know that $f_2(x) \ge 0$ for $0 \le x \le x^*$, so we have proved the inequality $(rw - hD) \bphi_w - \la \bphi \le 0$ on this subregion.

We next demonstrate the inequality $D_0(w) \le D_j(w)$ for $w^* \le w \le {hb \over r+h}$.  We know that the inequality holds with equality at $w = {hb \over r+h}$; thus, we only need to prove it for $w^* \le w < {hb \over r+h}$.  By setting $x = f_j(w)$ and simplifying, this inequality becomes
$$
1 - {r+h \over r} \, x^{{\la \over r} - 1}  + {h \over r} \, x^{{\la \over r}} \le 0
$$
for $x^* \le x < 1$, which holds because the left side is $f_3$ of Lemma 3.4 when we set $a = {\la \over r}$ and $c = {\la \over r+h}$.

We have shown that $\bphi$ in (3.32) satisfies the conditions of Lemma 3.8.  Thus, $\bphi$ is the maximum probability of reaching the bequest goal before ruining.  \qed

There exist remarks and a corollary to Proposition 3.16 for the case $\la > r$ that are parallel to those for $\la < r$, namely, Remarks 3.11 and 3.12 and Corollary 3.15. The only change is in definition of the regions on which the optimal behaviors of waiting and buying full insurance occur.  In the interest of space, we omit those remarks and corollary.

In the following theorem, we provide the reader with a summary of the maximum probability of reaching the bequest goal without ruining for the problem in this section.  We also provide a diagram describing the optimal insurance purchasing strategy in the case for which $\la > r$; see Figure 3.1.  A similar diagram applies when $\la \le r$; in that case $w^*$ would be 0.  In the interest of space, we do not include the corresponding summary theory for expected wealth at death.

\thm{3.17} {Divide the region $\bDo = \{ (w, D): 0 \le w \le \bar w(D), \; D \ge 0\}$ into the following four regions:
$$
\eqalign{
&\bDo_0 = \{ (w, D): 0 \le w \le {hD \over r}, \;  D \ge b \}, \cr
&\bDo_a = \{ (w, D): b - D \le w \le {hD \over r}, \;  {rb \over r+ h} < D < b \}, \cr
&\bDo^w_b = \{ (w, D): 0 \le D \le D_j(w), \;  0 \le w \le {hb \over r+h} \}, \cr
&\bDo^j_b = \bDo - \left( \bDo_0 \cup \bDo_a \cup \bDo^w_b \right).
}
$$
Then, the maximum probability of reaching the bequest goal without ruining is given by
$$
\bphi(w, D) =
\cases{ 1 - \left( {hD - rw \over hD} \right)^{{\la \over r}}, &if $(w, D) \in \bDo_0$, \cr
1 - \left( {(r+h)D - rb \over hb} \right)^{{\la \over r+h}}  \left( {hD - rw \over (r+h)D - rb} \right)^{{\la \over r}}, &if $(w, D) \in \bDo_a$, \cr
\left( {rw - hD \over h \left( {rb \over r+h} - D \right)} \right)^{{\la \over r}}, &if $(w, D) \in \bDo^w_b$, \cr
1 - \left( {hb - (r+h)w \over hb} \right)^{{\la \over r+h}}, &if $(w, D) \in \bDo^j_b$.}
$$
The optimal insurance purchasing strategy is as follows:
\item{$(a)$} If $(w, D) \in \bDo_0$, then do not buy any additional life insurance.
\item{$(b)$} If $(w, D) \in \bDo_a$, then buy additional life insurance only when wealth reaches $b - D$, after which continually buy life additional insurance to ensure that the sum of wealth and death benefit equals $b$.
\item{$(c)$} If $(w, D) \in \bDo^w_b$, then do not buy additional life insurance until wealth reaches the safe level ${hb \over r+h}$, at which time buy additional life insurance of ${rb \over r+h} - D$.
\item{$(d)$} If $(w, D) \in \bDo^j_b$, then immediately buy additional life insurance of $b - (w+D)$ and thereafter buy life insurance to ensure that the sum of wealth and death benefit equals $b$.}

\sect{4. Summary and Conclusions}

We determined the optimal strategies for purchasing life insurance in order to maximize the probability of reaching a bequest goal $b$.  When insurance is purchased by a single premium, as in Section 2, with or without a cash value, it is  optimal to wait until wealth reaches the safe level to buy additional life insurance.  When there is a cash value, then for wealth low enough, the individual will surrender {\it all} her life insurance and thereafter follow the optimal strategy from the no-cash-value case for buying life insurance, that is, wait until wealth reaches the safe level.  If the individual has no life insurance initially, then the existence of cash value is immaterial to her because she will not buy any life insurance until her wealth reaches the safe level.

When insurance is purchased by a premium payable continuously, we considered two cases in Section 3:  (1) instantaneous term life insurance, and (2) irreversible whole life insurance.  Arguably, whole life insurance is not irreversible due to standard nonforfeiture laws; however, in our time-homogeneous case, no reserve develops, so there is no cash value associated with the whole life insurance policy.  Thus, either we have the extreme in which we allow the individual complete freedom as to the amount of life insurance she purchases at any  time, as in case (1), or we have the extreme in which we do not allow the individual to reverse her purchase of life insurance once she decides to buy it, as in case (2).

For instantaneous term life insurance, which we considered in Section 3.1, if the force of mortality $\la$ is no greater than the return on the riskless asset $r$, then the individual does not buy any additional insurance until her wealth reaches the safe level.  If the force of mortality $\la$ is greater than the return on the riskless asset $r$, and if wealth is large enough, she will follow this same strategy of waiting; however, if wealth $w$ is small enough, she will buy full insurance ($D = b - w$) for the remainder of her life.  

For irreversible whole life insurance, the solution is more complicated, but if the initial death benefit is less than $b - w$, then we saw the same kind of optimal life insurance purchasing strategies there: if wealth is close enough to the safe level, then the individual will wait; if wealth is close enough to $b- D$, then the individual will buy so-called full insurance.  It is interesting that if the individual in Section 3.2 has no life insurance initially, then her optimal insurance purchasing strategy is {\it identical} to the corresponding one in Section 3.1 (depending on $\la$ versus $r$ and her initial wealth) because the optimal life insurance purchasing strategies in Section 3.1 are non-decreasing.

In future work, we will consider various extensions of the set up in Section 3, in which the life insurance premium is payable continuously.  We will (1) allow the force of mortality to change with time, as well as the cost of insurance; thus, for whole life insurance, the policy will develop a cash value; (2) maximize expected wealth at death, possibility limited by a given amount to prevent unrealistic life insurance purchasing strategies; and (3) assume that the individual also consumes from her wealth and wishes to maximize the probability of reaching a bequest goal without experiencing bankruptcy, possibly including life annuities in the financial market to cover some or all of her expenses.   We anticipate that the solutions to these problems will be important because they will be directly applicable to financial planning.

\sect{Appendix A. Proof of Lemma 3.4}

{\bf Proof of (a).}  Observe that $f_1(0) = f_1(1) = 0$.  Also,
$$
f'_1(x) = a \, x^{a-1} - c \, (1-x)^{c-1},
$$
and note that $f'_1(0) = -c < 0$ and $\lim_{x \to 1} f'_1(x) = -\infty$.  Thus, $f_1$ has an odd number of zeros in the interior $(0, 1)$, say, $2k - 1$, for some $k = 1, 2, \dots$.  This fact implies that $f'_1$ has $2k$ zeros; thus, to show that $k = 1$, it is enough to show that $f'_1$ has at most two zeros in $(0, 1)$.  The zeros of $f'_1$ are those points $x$ that solve
$$
x^{a-1} (1-x)^{1-c} = {c \over a}.
$$
So, if we define $g$ by
$$
g(x) = x^{a-1} (1-x)^{1-c} - {c \over a},
$$
then to show that $f'_1$ has at most two zeros in $(0, 1)$, it is enough to show that $g'$ has one zero in $(0, 1)$ because $g(0) = g(1) = - {c \over a} < 0$.  This result follows from
$$
g'(x) = x^{a-2} (1-x)^{-c} [(a-1) - (a-c)x],
$$
which has a unique zero at $x = {a-1 \over a-c} \in (0, 1)$.  Thus, we have proved that $f_1$ has a unique zero in $(0, 1)$.

{\bf Proof of (b).}  Observe that $f_2(0) = 0$ and $\lim_{x \to 1} f_2(x) = - \infty$. Also,
$$
f'_2(x) = {c \over a} \, (1 - x)^{c-2} \left[ (a-1) - (a-c) x \right];
$$
thus, $f_2$ increases on $\left[0, {a - 1 \over a - c} \right)$ and decreases on $\left( {a - 1 \over a - c}, 1 \right)$.  To show that $f_2$ is non-negative on $[0, x^*)$, it is, therefore, enough to show that $f_2(x^*) \ge 0$.

To this end, recall that $(1 - x)^c \le 1 - cx$ because the left side of this inequality is concave in $x$ (so lies below its tangents) and the right side is the tangent of $(1-x)^c$ at $x = 0$.  From part (a), we know that $1 - (x^*)^a = (1-x^*)^c$; thus, we conclude that $c x^* \le (x^*)^a$.  Inequality $f_2(x^*) \ge 0$ is equivalent to
$$
{a(1 - x^*) \over a(1-x^*) + cx^*} \ge (1 - x^*)^c,
\eqno(A.1)
$$
which will follow if we show the stronger inequality
$$
{a(1 - x^*) \over a(1-x^*) + (x^*)^a} \ge 1 - (x^*)^a,
\eqno(A.2)
$$
in which we use $1 - (x^*)^a = (1-x^*)^c$ and $c x^* \le (x^*)^a$.  Inequality (A.2) is equivalent to
$$
(x^*)^a + a(1-x^*) - 1 \ge 0,
$$
which holds on $[0, 1)$ because the left side decreases with respect to $x^*$ and equals $0$ if $x^* = 1$.  Thus, we have proved that $f_2$ is non-negative on $[0, x^*)$.

{\bf Proof of (c).}  Observe that $f_3(0) = 1$ and $f_3(1) = 0$.  Also,
$$
f'_3(x) = {a \over c} \, x^{a - 2} \left[ -(a-1) + (a-c) x \right];
$$
thus, $f_3$ decreases on $\left[0, {a - 1 \over a - c} \right)$ and increases on $\left( {a - 1 \over a - c}, 1 \right)$.  To show that $f_3$ is non-positive on $[x^*, 1]$, it is, therefore, enough to show that $f_3(x^*) \le 0$.  Inequality $f_3(x^*) \le 0$ is equivalent to
$$
{cx^* \over  a(1 - x^*) + cx^*} \le (x^*)^a,
$$
which is equivalent to inequality (A.1) because $(x^*)^a = 1 - (1-x^*)^c$.  Thus, we have proved that $f_3$ is non-positive on $[x^*, 1]$.

\sect{Appendix B. Proof of Lemma 3.13}

{\bf Proof of a.} $D_j(w) \le {rw \over h}$ if and only if
$$
{w - {hb \over r+ h} f_j(w) \over 1 - f_j(w)} \le w,
$$
which is equivalent to $f_j(w) = 0$, which only occurs when $w = 0$, or $f_j(w) \ne 0$ and $w \le {hb \over r+h}$.  It follows that $D_j(w) \le {rw \over h}$ holds for $0 \le w \le {hb \over r+h}$, with equality only at the endpoints.  This result confirms our hypothesis that the line $rw = hD$ lies in the jump region.

\medskip

To prove parts (b) and (c), we will use the fact that $D'_j(w)$ is proportional to the following:
$$
D'_j(w) \propto 1 - {h \over r+h} f_j(w) - {r \over r+h} f_j(w)^{1 - {\la \over r}}.
$$

{\bf Proof of b.} If $\la \le r$, then $D_j(w)$ is increasing on $\left[0, {hb \over r+h} \right]$ if and only if 
$$
1 - {h \over r+h} \,  x - {r \over r+h} \, x^{1 - {\la \over r}} \ge 0,
$$
for $0 \le x \le 1$.  This inequality holds because the left side equals 0 when $x = 1$, and the left side decreases on $[0, 1]$.

{\bf Proof of c.}  If $\la > r$, first note that $D_j(w)$ has a unique zero at $w = w^*$ because $D_j(w) = 0$ if and only if the expression in (3.9) equals 0.  Because $D'_j(0) < 0$, we conclude that $D_j(w) \le 0$ for $0 \le w \le w^*$.  Also, note that $f_j(w)$ increases on $\left[0, {hb \over r+h} \right]$ and equals $x^*$ when $w = w^*$.  It follows that $D_j(w)$ is increasing on $\left[w^*, {hb \over r+h} \right]$ if and only if 
$$
1 - {h \over r+h} \, x - {r \over r+h} \, x^{1 - {\la \over r}} \ge 0,
$$
for $x^* \le x \le 1$, which is equivalent to $f_3(x) \le 0$ for $x^* \le x \le 1$, which we know is true from Lemma 3.4.

\bigskip

\centerline{\bf Acknowledgments} \medskip  The first and third authors thank the Committee for Knowledge Extension and Research of the Society of Actuaries for financially supporting this work.  Additionally, research of the first author is supported in part by the National Science Foundation under grants DMS-0955463, DMS-0906257, and the Susan M. Smith Professorship of Actuarial Mathematics. Research of the third author is supported in part by the Cecil J. and Ethel M. Nesbitt Professorship of Actuarial Mathematics.

\sect{References}

\noindent \hangindent 20 pt Bayraktar, Erhan and Virginia R. Young (2013), Life insurance purchasing to maximize utility of household consumption, {\it North American Actuarial Journal}, 17 (2): 114-135.

\smallskip \noindent \hangindent 20 pt Browne, Sid (1997), Survival and growth with a liability: optimal portfolio strategies in continuous time, {\it Mathematics of Operations Research}, 22 (2): 468-493.

\smallskip \noindent \hangindent 20 pt Browne, Sid (1999a), Beating a moving target: optimal portfolio strategies for outperforming a stochastic benchmark, {\it Finance and Stochastics}, 3 (3): 275-294.

\smallskip \noindent \hangindent 20 pt Browne, Sid (1999b), Reaching goals by a deadline: digital options and continuous-time active portfolio management, {\it Advances in Applied Probability}, 31 (2): 551-577.



\smallskip \noindent \hangindent 20 pt Dubins, Lester E. and Leonard J. Savage (1965, 1976), {\it How to Gamble if You Must: Inequalities for Stochastic Processes}, 1965 edition McGraw-Hill, New York. 1976 edition Dover, New York.



\smallskip \noindent \hangindent 20 pt Kulldorff, Martin (1993), Optimal control of favorable games with a time limit, {\it SIAM Journal on Control and Optimization}, 31 (1): 52-69.


\smallskip \noindent \hangindent 20 pt Milevsky, Moshe A., Kristen S. Moore, and Virginia R. Young (2006), Asset allocation and annuity-purchase strategies to minimize the probability of financial ruin, {\it Mathematical Finance}, 16 (4): 647-671. 


\smallskip \noindent \hangindent 20 pt Pestien, Victor C. and William D. Sudderth (1985), How to control a diffusion to a goal, {\it Mathematics of Operations Research}, 10 (4): 599-611.

\smallskip \noindent \hangindent 20 pt Richard, Scott F. (1975), Optimal consumption, portfolio and life insurance rules for an uncertain lived individual in a continuous time model, {\it Journal of Financial Economics}, 2 (2): 187-203.

\smallskip \noindent \hangindent 20 pt Sudderth, William D. and Ananda Weerasinghe (1989), Controlling a process to a goal in finite time, {\it Mathematics of Operations Research}, 14 (3): 400-409.

\smallskip \noindent \hangindent 20 pt Wang, Ting and Virginia R. Young (2012a), Optimal commutable annuities to minimize the probability of lifetime ruin, {\it Insurance: Mathematics and Economics}, 50 (1): 200-216.

\smallskip \noindent \hangindent 20 pt Wang, Ting and Virginia R. Young (2012b), Maximizing the utility of consumption with commutable annuities, {\it Insurance: Mathematics and Economics}, 51 (2): 352-369.


 \bye